# A Deployment Process for Strategic Measurement Systems

Martin Kowalczyk, Henning Barthel, Jürgen Münch, Jens Heidrich and Adam Trendowicz

## Abstract

*Explicitly linking software-related activities to an organization's higher-level goals has been shown to be critical for organizational success. GQM$^+$Strategies[1] provides mechanisms for explicitly linking goals and strategies, based on goal-oriented strategic measurement systems. Deploying such strategic measurement systems in an organization is highly challenging. Experience has shown that a clear deployment strategy is needed for achieving sustainable success. In particular, an adequate deployment process as well as corresponding tool support can facilitate the deployment. This paper introduces the systematical GQM$^+$Strategies deployment process and gives an overview of GQM$^+$Strategies modelling and associated tool support. Additionally, it provides an overview of industrial applications and describes success factors and benefits for the usage of GQM$^+$Strategies.*

## 1. Introduction

Software is becoming increasingly a pervasive element in many business areas. Consequently, explicitly linking software activities to an organization's higher-level business goals is becoming increasingly important, because this alignment can improve organizational performance [8]. Furthermore, studies show that funding decisions on information technology investments are increasingly made on higher levels of company hierarchies, thus rather by managers than by technicians [5]. Based on the authors' experience, being able to show the contribution of software-related activities to the strategic objectives of an organization significantly helps in obtaining sufficient resources for critical activities. Hence, it becomes critical for software and IT departments to quantify their value contribution to the overall business goals in order to avoid irrational budget cuts and personnel reductions. Strategic measurement systems can support organizations in demonstrating the respective contributions of different organizational units to the overall business goals, as well as provide a means for success monitoring and evaluation.

GQM$^+$Strategies [2],[3] is an approach for clarifying and harmonizing goals and strategies across all levels of an organization and communicating business goals throughout an organization. Furthermore, the approach supports monitoring the success or failure of strategies and business goals [3]. One central result of the approach is the GQM$^+$Strategies Grid, which is a model of a strategic measurement system. It specifies goals and strategies across all levels of an organization, including the measurement program needed to monitor and control them.

Experience gained by the authors when developing such grids with an organization has shown that a clear deployment strategy is needed in order to achieve sustainable results. A systematic deployment process, which does not only address engineers on the lower levels of an organization but also managers on the top level, is essential for creating true commitment across all organizational levels and establishing a GQM$^+$Strategies Grid.

For modelling such a grid, corresponding tool support is needed to address the specific viewpoints and information needs of all stakeholders. Furthermore, a systematic approach to maintaining such a grid is needed. As an organization's goals and strategies change and mature, the GQM$^+$Strategies Grid needs to be adapted accordingly in order to reflect an up-to-date image of the organization. Only then a GQM⁺Strategies Grid can be used as an active means to manage goals and strategies by measuring their success and failures and initiating corresponding improvement programs.

---

[1] GQM$^+$Strategies is registered trademark No. 302008021763 at the German Patent and Trade Mark Office; international registration number IR992843.

This paper discusses the GQM⁺Strategies deployment process, including aspects of integration and maintenance, as well as tool support for modelling. Section 2 of this paper gives an overview of related work. Section 3 discusses GQM⁺Strategies modelling and introduces corresponding tool support. Section 4 provides insights into the GQM⁺Strategies deployment process that was developed in order to systematically support the application of the approach in industry. Section 5 sketches practical industry applications and highlights success factors and benefits for the implementation of strategic measurement systems using GQM⁺Strategies. Section 6 summarizes the paper and briefly discusses future work.

## 2. Related work

The discussion of different concepts is often hampered by semantically implicit definitions and understandings. In the context of strategic management, our experience has shown that a plethora of terms is in use. For example, for the terms goals and strategies many different conceptions exist, which tends to create misunderstanding in their usage. To improve the semantical clearness of the GQM⁺Strategies approach, we would therefore like to start with a short discussion and comparison of GQM⁺Strategies and the Business Motivation Model (BMM) [10]. The BMM uses the notions of *Ends* and *Means* for differentiating aspects of business that are aspired from the actual implementation or action plans. Ends are refined for different abstractions level into *Vision*, *Goal,* and *Objective*. The same accounts for Means, which are refined into the concepts of *Mission, Strategy,* and *Tactic*. GQM⁺Strategies uses the notions of *Goals* and *Strategies* on all the different levels of an organization to describe aspirations and the courses of actions needed to achieve the aspired states. *Figure 1* shows a simple mapping of these concepts. The BMM does not represent the shown linkages between the ends and means concepts explicitly, but states that they are feasible. In contrast, linking goals and strategies across multiple organizational levels is essential for the GQM⁺Strategies approach, as this creates traceable connections between goals, strategies, and measurement models.

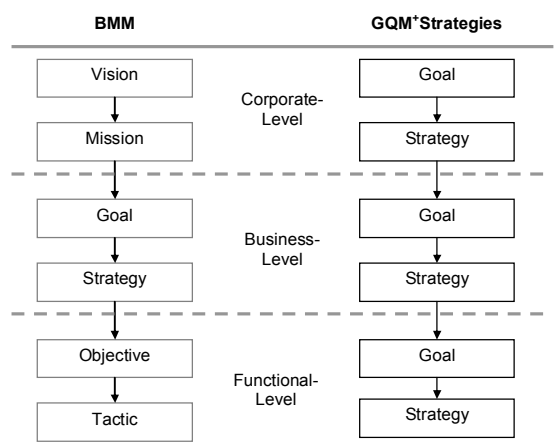

*Figure 1: Simple mapping of BMM and GQM⁺Strategies*

Often, top-level measurement systems are defined without creating such explicit linkages. A popular approach is the Balanced Scorecard (BSC) [7], which defines four perspectives, typically financial, customer, internal business processes, and learning or growth, and links objectives from these perspectives to measures. Strategy maps are used to link strategies to respective goals and perspectives. In this scenario, misalignment between different organizational levels is possible [4]. Becker and Bostelman [4] see two causes of misalignment: (1) project data that does not address organizational goals and (2) organizational goals that are not operationalized through processes and metrics at the project level. Their approach is to embed a GQM structure within each of the four BSC perspectives. The GQM approach [1] provides a method for defining goals, refining them into questions and finally data to be collected, and then analyzing and interpreting them. However, it does not explicitly link its measurement goals to organizational goals and strategies at different levels of an organization. The GQM$^+$Strategies approach addresses these issues [2].

Approaches like CobiT [6] and ITIL [11] strictly focus on traditional or service-oriented IT governance. Thus, they provide connections between sets of goals and attributes of the IT infrastructure. For example, CobiT uses a fixed linkage structure between outcome measures and performance indicators on the business, IT, process, and activity levels. Although these approaches recognize the need to link organizational goals and measures, they focus only on a very small subset of an organization. With increasing relevance of software throughout all aspects of business, approaches are needed that can capture these business scenarios more specifically. GQM$^+$Strategies aims at providing these capabilities. It allows for modelling of goals and strategies at different levels of an organization and addresses problems like adequate visualization to support decision-making.

## 3. GQM$^+$Strategies modelling

The GQM$^+$Strategies approach is an extension of the GQM approach for goal-oriented measurement and provides capabilities for linking and communicating goals and strategies across different levels of an organization. Thus, it allows monitoring the success or failure of strategies and organizational goals using a strategic measurement system. This section will discuss the modelling of strategic measurement systems using the GQM$^+$Strategies approach. For this purpose, we will introduce the GQM$^+$Strategies conceptual model. Furthermore, we will present the currently available tool support for modelling and visualizing GQM$^+$Strategies Grids.

### 3.1. GQM$^+$Strategies modelling basics

Modelling strategic measurement systems that link and control organizational goals and strategies across multiple organizational levels requires concepts for adequately representing organizational goals and strategies as well as concepts that support the definition of measurement models. The GQM$^+$Strategies conceptual model (see *Figure 2*) addresses both aspects.

Goal$^+$Strategies elements (see left side of *Figure 2*) provide the capability to define linked sequences of goals and associated strategies. Strategies describe a planned and goal-oriented course of actions for achieving the defined goals at the respective organizational level. The conceptual model allows multiple goal levels and permits deriving multiple strategies for each of these goal levels. A *goal* may be realized by a set of *strategies*, which may in turn lead to a set of goals. Additionally, Goal$^+$Strategies elements provide the capability to capture the underlying rationales for the defined goals, strategies, and their linkages using *context factors* and *assumptions*. Context factors are environmental variables that represent the organizational environment and affect the kind of models and data that can be used. Assumptions are estimated unknowns that can affect the interpretation of measurement data and, consequently, the interpretation of related goals and strategies.

The entire model provides an organization with a mechanism not only for modelling goals and strategies but also for defining measurement consistent with high-level organizational goals and for interpreting and rolling up the resulting measurement data at each level. For this purpose, the GQM approach is used, which constitutes the main element within the GQM graphs representing the measurement part of the conceptual GQM+Strategies model (see right side of *Figure 2*). A *GQM graph* consists of a GQM goal (that measures a GQM+Strategies element), corresponding questions, metrics, and additional interpretation models. At each goal level, such a GQM graph is modelled in order to measure the achievement of the defined goal in combination with the chosen strategy. Accordingly, the definition of a complete measurement plan includes the definition of GQM measurement goals, the derivation of questions and metrics, as well as the definition of an interpretation model that determines whether the measurement goal has been achieved.

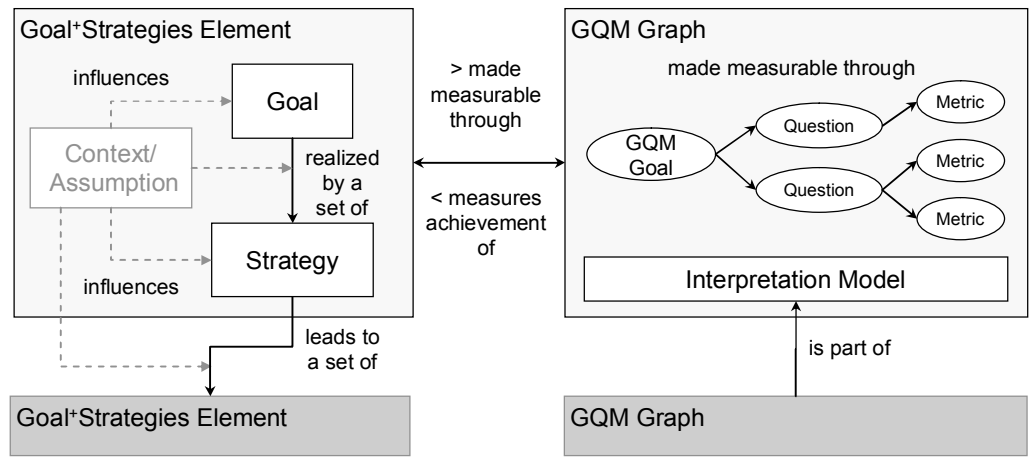

*Figure 2: GQM+Strategies conceptual model*

Goal+Strategies elements and associated GQM graphs are the components that are used to model complex strategic measurement systems. Such a measurement system, the *GQM+Strategies Grid,* specifies goals and strategies across all levels of an organization, including the GQM models needed to monitor and control them. Modelling a grid helps to make goals and strategies explicit for a whole organization or selected parts thereof and provides a clear linkage to all of the organization's measurement initiatives. In consequence, each organizational unit that is involved in the modelling scope profits from clearly defined goals and strategies as well as a clear link to upper-level goals to which they contribute. Additionally, these organizational units are able to define clearly linked goals and strategies for all operational activities within the respective unit. One of the greatest benefits is that all of the organization's measurement initiatives that are related to these goals and strategies are integrated into one grid. This creates transparency and is the basis for measurement efficiency, as there is a central measurement system that integrates all organizational measurement initiatives.

Depending on the scope of the application, different kinds of modelling support are adequate. If only a small scope needs to be addressed, more traditional tools like a whiteboard and a presentation for the dissemination of results might suffice. *Figure 3* shows the modelling results for excerpts of a sample grid that was created based on a hypothetical organization, but built upon real project experience gained over many years of industrial work. The grid encompasses three organizational levels with one goal and strategy per level. In addition, the associated measurement models are represented, which measure the success or failure of goals and strategies. In most industry cases, the resulting grid consists of a lot more elements (for an example see [9]). There are usually multiple goals and strategies per organizational level that need to be modelled, and the measurement models are often more complex. This creates quite challenging requirements with respect to convenient ways of modelling, representing, and communicating a GQM+Strategies Grid. Consequently, there is a need for more elaborated ways of modelling and visualizing GQM+Strategies Grids. The following section will introduce the currently available tool support for modelling and visualizing complex GQM+Strategies Grids.

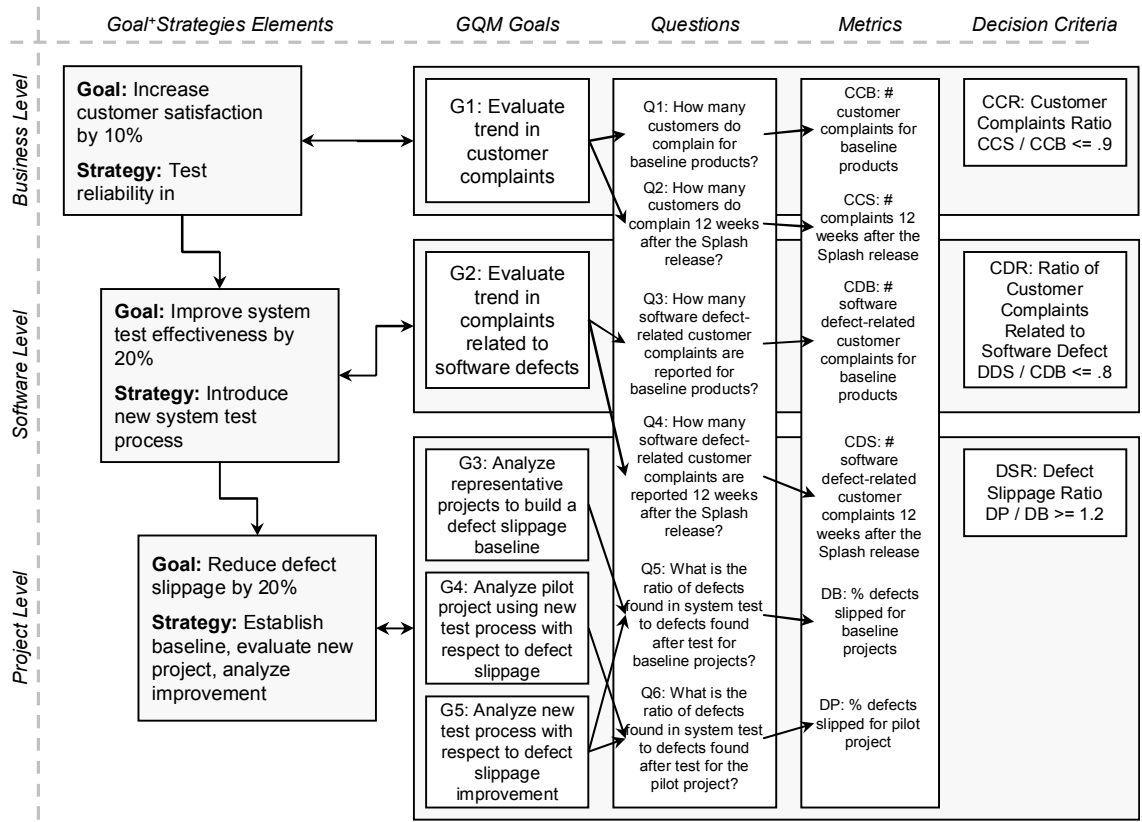

Figure 3: Exemplary GQM+Strategies Grid

## 3.2. Tool support for GQM+Strategies modelling

GQM+Strategies Grids with multiple goals and strategies across several organizational levels within an organization can get very large and complex. To achieve a convenient way of modelling and representing such grids, templates for goals, strategies, context factors, assumptions, and GQM models are used. These templates guide users in the modelling process and support them in explicitly capturing and defining all measurement entities at the necessary level of detail. However, experiences gained with larger grids show that understanding the effects of any changes in the context of a grid's entire set of goals and strategies becomes quite difficult when relying solely on these templates. Effective decision-making requires efficient ways of analyzing goal and strategy interrelationships and changes within the grid. Therefore, appropriate visualization techniques are necessary that provide an overview of the overall grid and offer dedicated abstraction, interaction, and drill-down capabilities. We used the InSEViz framework, developed at Fraunhofer IESE, to realize a visual 2D grid representation. *Figure 4* shows a sample GQM+Strategies Grid visualization.

Node-link diagrams representing connections as edges between vertices are used for visualizing the relationships between different objectives at various organizational levels. The goals and strategies as well as the context factors and assumptions attached to them are visualized using graphical node elements such as text, geometric shapes, images, shadows, and colour fills. The same approach is used for visualizing the GQM models. All nodes as well as geometrical edges representing the links between them are automatically layouted, reflecting the overall structure of the grid. This supports the user in better understanding the relationships between goals and strategies across all levels of the organization, which improves the traceability and maintainability of the GQM[+]Strategies Grid.

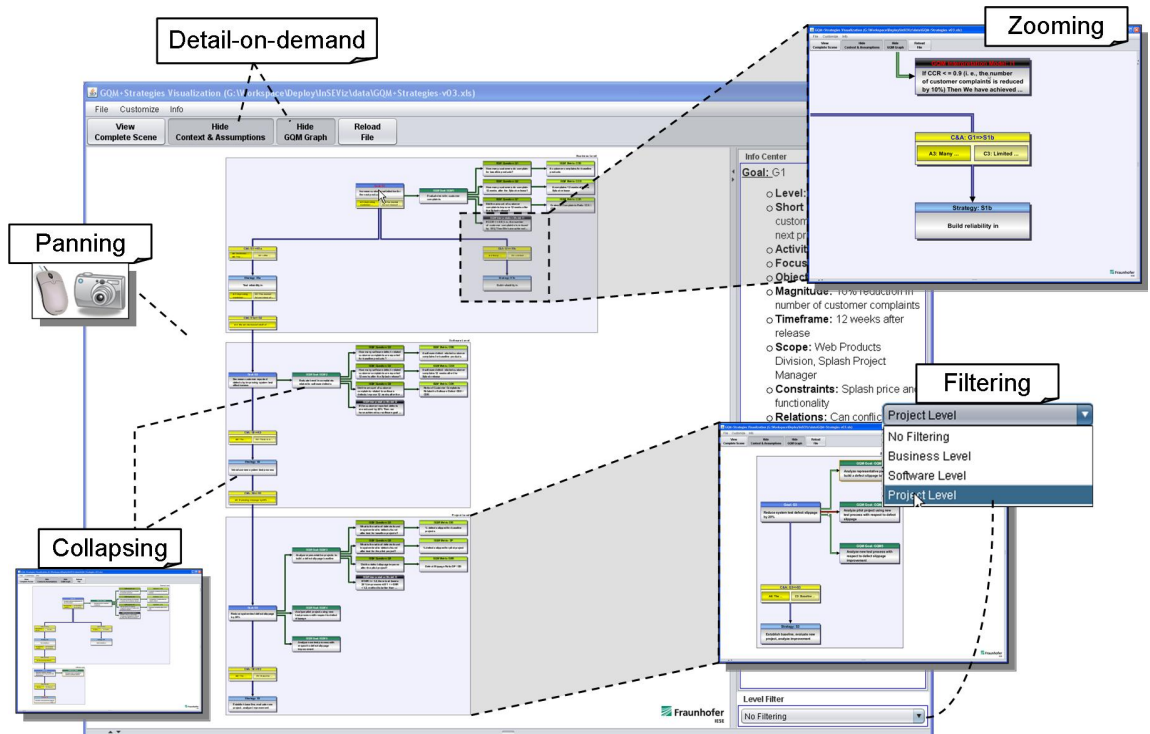

*Figure 4: Exemplary visualization of a GQM[+]Strategies Grid*

However, a large number of nodes and links increase the overall visual, and thus the cognitive, complexity of a grid. In order to reduce this complexity, a filtering approach that implements concept- and role-oriented views is used. For instance, according to his role, a user can switch between different views showing the business goals and strategies or just the corresponding GQM graph. The view-based approach supports grid users with analysis and reasoning tasks and is improved further by the following interaction and filtering operators (see *Figure 4*):

- Panning and zooming: A virtual camera within the geometric space of the visualization treats the data display as a camera, enabling the user to interactively pan and zoom within the scene to explore the grid. Moving and zooming the camera accordingly changes the viewpoint seen by the user.
- Collapsing: The user can collapse single elements or edges to hide all the subsequent details and relations on lower levels.
- Detail-on-demand: Dependent on the zoom level, different levels of details are displayed to the user. Furthermore, the user can select whether to display or hide the GQM graphs or the context and assumption information for goals and strategies.
- Filtering: The framework uses several mechanisms for filtering out irrelevant data depending on the user's role. For instance, higher levels of the organization may be filtered out for users on lower levels.

These pre-built interaction and filtering operators are independently customizable by the user for each view and thus help to reduce complexity during modelling and usage of GQM[+]Strategies Grids.

# 4. The GQM+Strategies Process

A comprehensive and systematic deployment strategy for the introduction of the GQM+Strategies approach into industry organizations is of great importance. An effective deployment strategy has to address all relevant stakeholders and has to provide an efficient way for introducing the GQM+Strategies approach. The previous section described the modelling concepts and tool support available for modelling, visualization, and communication of the grid within an organization. This section will focus on the GQM+Strategies Process, which provides systematic support on how to set up and maintain a strategic measurement system using the GQM+Strategies approach.

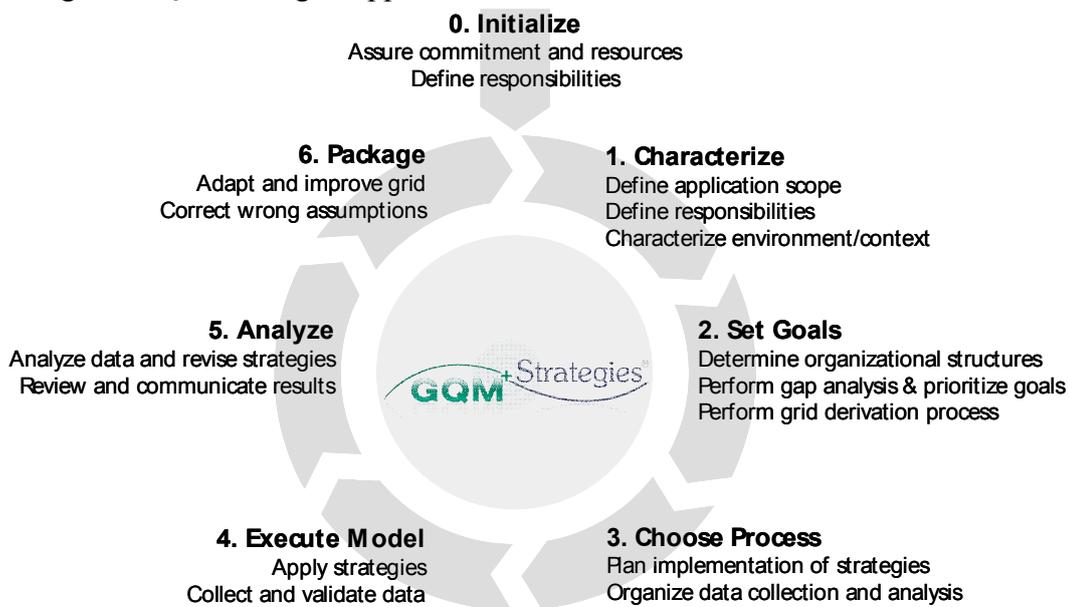

Figure 5: The GQM+Strategies Process based on QIP

The GQM+Strategies Process was developed in joint collaboration between the Fraunhofer Institute for Experimental Software Engineering (IESE) and the Fraunhofer Center for Experimental Software Engineering (CESE). It provides support for deploying the GQM+Strategies approach to industry. Following this process provides a systematic way for linking organizational goals, creating a custom-tailored, goal-oriented strategic measurement system, and maintaining the strategic measurement system. The GQM+Strategies Process consists of seven process steps and was created on the conceptual basis of the Quality Improvement Paradigm (QIP) [1]. In the following, we will first give an overview of these seven process steps with a focus on the GQM+Strategies Grid derivation process and then address aspects of process integration and grid maintenance.

In accordance with the QIP cycle, we defined six basic process steps and added one additional step for the initialization of the GQM+Strategies approach (see *Figure 5*). Thus, the GQM+Strategies Process encompasses the steps Initialize, Characterize, Set Goals, Choose Process, Execute Model, Analyze, and Package. The GQM+Strategies Process with its seven steps aims at supporting the creation of custom-tailored GQM+Strategies Grids and associated measurement systems.

*Initialize* creates the conditions required for successful application of the GQM+Strategies approach by assuring organizational commitment and resources for the application of GQM+Strategies. Furthermore, the GQM+Strategies application process is planned, responsibilities are defined, and GQM+Strategies training is provided to all persons who will be involved during deployment.

*Characterize* defines the scope and environment of the GQM[+]Strategies application. Scope defines those parts of an organization that are important for the GQM[+]Strategies application, and environment describes the context and underlying assumptions with respect to the defined scope.

*Set Goals* represents the conceptual core of the GQM[+]Strategies approach, as in this step, the GQM[+]Strategies Grid is created. For the creation of the grid, the relevant organizational levels have to be defined and already existing assets (i.e., goals, strategies, measures, etc.) have to be determined. Based on this information, a gap analysis is performed. The gap analysis identifies missing elements or ambiguities in the current set of assets. These could be missing goals, strategies, or measures, or unresolved conflicts of alignment in the set of assets. After the gap analysis, all relevant goals are defined and prioritized. The obtained set of goals is then refined into the GQM[+]Strategies Grid, by deriving strategies linking them to goals at the respective organizational levels. During this derivation process, organizational levels are described and goals and strategies are modelled in a formalized way. This supports understanding and communication of goals and strategies throughout the organizations. Furthermore, capturing the rationale for the defined goals and strategies is achieved by documenting respective context factors and assumptions. Additionally, during *Set Goals*, the derived goals and strategies are made measurable by defining corresponding GQM measurement goals and using the traditional GQM approach [1] to refine these goals into questions and metrics. GQM[+]Strategies measures goal achievement and evaluates the effectiveness of the performed strategies. This is realized by setting up so-called interpretation models, which specify how to evaluate goal achievement.

*Choose Process* makes measurement operational. Activities that are performed during this step are the creation of a measurement plan (i.e., what will be collected by whom, when and how often, and how) and the preparation of procedures and tools for data collection, analysis, and interpretation.

During *Execute Model*, the actual data collection and validation is performed according to the defined measurement plan. For the purpose of data collection, some kind of data collection mechanism has to be implemented.

*Analyze* refers to data analysis and visualization. This includes applying the defined GQM[+]Strategies interpretation models to the measurement data collected and making statements about goal achievement. Analysis may lead to the identification of new gaps in the GQM[+]Strategies Grid and thus can trigger modification suggestions and maintenance actions for the existing grid.

*Package* provides the activities that are necessary for keeping the measurement system up to date with respect to the current organizational goals and strategies and is therefore the most important step for the maintenance of a GQM[+]Strategies measurement system. All needed modifications are documented and required changes are performed. The results are packaged into the organization's experience base.

In order to provide flexibility and an up-to-date picture of the organization, the GQM[+]Strategies Process incorporates the possibility to perform maintenance activities during several steps within this process as described in the following.

(1) In subsequent iterations of the whole GQM[+]Strategies process, the initialization step is revisited to check whether maintenance actions are needed.
(2) Revision due to changes in scope and environment can be performed within the characterization step.
(3) During the set goals step, elements of the GQM[+]Strategies Grid can be revised and changed according to the modelling procedures of GQM[+]Strategies.

(4) Finally, maintenance actions that are needed due to the analysis results of the current iteration of the GQM$^+$Strategies Process are performed during the packaging step, as described above.

The iteration intervals for application and maintenance of GQM$^+$Strategies strongly depend on organizational planning processes, which typically are planning, plan implementation, controlling and plan adaptation. The GQM$^+$Strategies Process needs to be integrated with these organizational processes, in order to be performed effectively and to unfold its benefits. Thus the application of the GQM$^+$Strategies approach and maintenance of the grid have to be performed in synchronization with the iteration intervals of organizational processes. For example, if the goals and strategies are revised every two years on the management level of an organization, then the GQM$^+$Strategies Process should be revisited during the same time interval and parts of the GQM$^+$Strategies Grid on the management level would be updated accordingly. If planning and revision of goals and strategies on the software level of this organization are done more frequently, e.g., on a yearly basis, the GQM$^+$Strategies Process should be applied more frequently for this level.

Deploying a well-integrated GQM$^+$Strategies Process in a systematic manner is essential for the success of the GQM$^+$Strategies application.

## 5. Applications, success factors and benefits

In this section, we give an overview of a number of applications where the GQM$^+$Strategies approach was used for introducing strategic measurement systems for different, organization-specific purposes. Additionally, we discuss success factors and benefits for the application of GQM$^+$Strategies.

### 5.1. Applications

*Application #1* was conducted in the context of a European telecommunication company. They used the approach to drive strategic improvement programs and to support the paradigm shift towards purpose-driven metrics. When the approach was applied, it became clear that easy-to-use templates had to be developed for documenting a GQM$^+$Strategies Grid and that the connection to existing standards needs to be clarified. When deploying the method and the resulting grid to the company, it was important to address operational/strategic planning including, for instance, the sequence and scheduling of different strategies. Moreover, the relationships between the different goals on different organizational levels needed to be clarified in terms of conflicting or supporting goals.

*Application #2* was conducted in the context of a European automotive supplier. Their intention was to support CMMI Measurement and Analysis, which was enforced by top-level management. They used the approach for harmonizing strategies and goals across different units and for defining corresponding management objectives. The company developed integrated tool support for GQM$^+$Strategies, on top of an existing GQM infrastructure. This includes mechanisms for storing and analyzing the evolution of all information of goals and strategies. Creating this transparency helped them to overcome mistrust and to define goals at the interfaces with collaborating organizations/units. It also became clear that resource needs can be justified in terms of the goals they are contributing to, and the consequences of budget cuts in terms of goal attainment could be illustrated. The company distinguished between different types of goal inheritance between different organizational levels, such as goals inherited identically from one level to another or goals inherited with different target values only, or goals refined by different sub-goals.

*Application #3* was conducted in the context of an Asian insurance company from the information systems domain. Their main motivation for applying the method was to align their strategies and goals for a new business domain and to quantitatively measure whether they would be able to achieve these goals. Moreover, they intended to clarify the goals and strategies of different organizational units involved and to define an IT strategy for the new domain. Finally, there was a need for a systematic way for synchronizing organizational goals and strategies with goals specified in individual projects. The method helped the company to identify gaps in the alignment of its goals, strategies, and measurement data, and to fill these gaps. We observed that for getting accepted in such an industrial setting, the GQM$^+$Strategies Grid needs to be documented in an easy-to-understand, easy-to-maintain, and easy-to-exchange manner. Tool support was needed for dealing effectively with the complex grid.

*Application #4* was conducted in the context of a joint research project sponsored by the German Federal Ministry of Education and Research. The project involved research and industry partners from various domains (e.g., logistics, industrial facilities). It aimed at developing a common software platform for supporting complex, dynamic business processes. The GQM$^+$Strategies approach helped project partners to clearly link their individual goals to project objectives, identify gaps and inconsistencies in project objectives, and understand each others' goals. The study showed that, in practice, an explicit and transparent way for modelling inter-related (supporting or contradicting) goals is needed for better understanding and communication.

*Application #5* was conducted in the context of an Asian systems engineering company from the safety-critical aerospace software domain [9]. The objective of the GQM$^+$Strategies application was to create transparency with respect to goals and strategies and derived measurement goals in the context of collaborations with suppliers. Additionally, the contributions of internal business units to top-level organizational goals should be made transparent and alignment should be secured. The application of GQM$^+$Strategies and the derivation of appropriate grids supported the creation of transparency regarding supplier collaboration and emphasized the contributions of internal business units to top-level organizational goals. The usage of the visualization tool was especially helpful for the early modelling steps, as it strongly enhanced understanding and communication.

## 5.2. Success factors and benefits

On the basis of these applications of the GQM$^+$Strategies approach, we could identify several properties of the approach and the resulting GQM$^+$Strategies Grids that are success factors for application in industry. This section gives an overview:

*(SF1) Tailorability:* GQM$^+$Strategies Grids can be custom-tailored to specific organizational needs, especially reflecting the underlying assumptions for choosing specific goals and strategies.

*(SF2) Traceability*: The GQM$^+$Strategies Grid defines links between different goals, strategies, and GQM models on different levels of the organization. For each of those elements, a decision maker can access the rationale (context and assumptions) for defining the element and the effects/relationships it has on other elements of the grid so that decisions can be made on a sound basis.

*(SF3) Understandability*: Goals and strategies need to be communicated within organizations. Therefore, the representation of the GQM$^+$Strategies Grid needs to be easy to understand by all stakeholders so that a clear plan of action can be derived. On the one hand, the GQM$^+$Strategies approach facilitates understandability by formalizing goals, strategies, as well as context and assumptions and thus reducing the fuzziness of these concepts. On the other hand, it integrates into existing organizational structures, goals, strategies, and measurement initiatives and usually preserves the natural semantics used within an organization.

*(SF4) Measurability*: Using the approach quantifies goals and their attainment so that deviations from target settings are easy to detect and countermeasures can be introduced effectively. Thus, the grid can be used as an active instrument for monitoring organizational goals and strategies.

*(SF5) Integration*: For carrying out the strategies of the grid and for monitoring and controlling the goals, the grid needs to be integrated into organizational and development processes. Moreover, the grid needs to be detailed to the operational level of the company, so that the defined strategies can be implemented.

Based on our experience from the applications, these success factors are crucial for the benefits that arise for organizations applying the GQM$^+$Strategies approach. These can be summarized as described in the following:

*(B1) Improved Communication*: Apparently the process of creating a GQM$^+$Strategies Grid is a vehicle for communication, as it requires different members of the organization to collaborate at multiple steps of the process. The resulting GQM$^+$Strategies Grid can improve communication with respect to organizational goals and strategies, as these are formalized, which allows for increased understandability. Additionally, the grid is tailored to the specific organizational environment and integrates into existing processes. These aspects facilitate acceptance and support different stakeholders in understanding each other and in coordinating decisions across different levels of an organization.

*(B2) Improved analysis, decision-making and controlling:* A GQM$^+$Strategies Grid that centrally integrates organizational measurement in a traceable manner with explicitly specified relationships, is the basis for sound analysis capabilities. Improved analysis capabilities support decision-making and controlling of goals and strategies. Additionally, the approach helps to increase the transparency of decision making, as a grid represents relevant organizational levels, provides traceability, and integrates into all relevant processes.

*(B3) Improved organizational learning*: Explicit documentation of rationales behind goals and strategies at each level in a grid allows for retrospective analysis of the grid. Thus, organizational goals and strategies are reflected upon within an iteration of the GQM$^+$Strategies process. Stakeholders may, for example, learn that they failed to consider some relevant context factors or that the assumptions they made were actually wrong. Based upon such retrospection, the grid can be tailored better to specific situations in the future.

## 6. Summary and Outlook

This paper presented the GQM$^+$Strategies Process, as a systematic approach for developing and deploying strategic measurement systems. First, we discussed the basics of GQM$^+$Strategies modelling as an important part of the deployment approach. In this context, we also shortly described the available tool support. We then focused on the GQM$^+$Strategies Process itself, which is designed to support organizations in introducing strategic measurement systems in a systematic manner.

The GQM$^+$Strategies Process consists of the seven steps Initialize, Characterize, Set Goals, Choose Process, Execute Model, Analyze, and Package. The purpose of the process is to systematically develop, deploy, and maintain a GQM$^+$Strategies Grid, which is the main component of a GQM$^+$Strategies measurement system. The GQM$^+$Strategies Grid specifies goals and strategies across all levels of an organization, including the GQM measurement models that are needed to monitor and control them. To ensure sustainability of organizational measurement, an approach for grid maintenance needed. We sketched the maintenance capabilities that are encompassed within the GQM$^+$Strategies approach. Interpretation models, context factors, and assumptions help to determine which parts of the model need to be changed/maintained and how to measure the effect of these changes.

However, it is also important that regular revisions are performed in order to check whether the model still reflects the organization properly. These revision cycles need to be synchronized with the planning processes at the different levels of the organization.

Furthermore, we presented some initial practical experience in the form of industrial applications and derived success factors for the usage of the approach in industry. Based on our experience, we additionally presented major benefits of the GQM$^+$Strategies approach.

Currently, further pilot studies are ongoing to obtain more insights into the deployment process. So far, most of the studies have been in the phase of setting up a GQM$^+$Strategies Grid, but not in the phase of actively using and maintaining the grid. A specific focus on theses phases will provide valuable insights into how to further improve the operationalization of GQM$^+$Strategies Grids. Additionally, we are developing an analysis framework that will allow us to evaluate the benefits of the approach in a more systematic manner. Regarding tool support, we are working on improving the visualization and modelling capabilities with mechanisms for feeding back information from the data analysis to the grid. This should help to capture and visualize the impact of empirical evidence on business goals.

## 7. Acknowledgments

This work was supported in part by the German Federal Ministry of Education and Research (BMBF) (Devise 01IS10013E, Optikon 01IS09049B and Quamoco 01IS08023C) and the Stiftung Rheinland-Pfalz für Innovation (Q-VISIT Project).